\documentclass[conference]{IEEEtran}

\usepackage{cite}
\usepackage{amsmath,amssymb,amsfonts,mathtools}
\usepackage{algorithmic}
\usepackage{graphicx}
\usepackage{textcomp}
\usepackage{xcolor}
\def\BibTeX{{\rm B\kern-.05em{\sc i\kern-.025em b}\kern-.08em
    T\kern-.1667em\lower.7ex\hbox{E}\kern-.125emX}}
\begin{document}

\title{Using analog scrambling circuits for automotive sensor integrity and authenticity}

\author{\IEEEauthorblockN{1\textsuperscript{st} Cristiano Pegoraro Chenet}
\IEEEauthorblockA{\textit{Control and Computer Engr. Dept.} \\
\textit{Politecnico di Torino}\\
Torino, Italy \\
cristiano.chenet@polito.it}
\and
\IEEEauthorblockN{2\textsuperscript{nd} Alessandro Savino}
\IEEEauthorblockA{\textit{Control and Computer Engr. Dept.} \\
\textit{Politecnico di Torino}\\
Torino, Italy \\
alessandro.savino@polito.it}
\and
\IEEEauthorblockN{3\textsuperscript{rd} Stefano di Carlo}
\IEEEauthorblockA{\textit{Control and Computer Engr. Dept.} \\
\textit{Politecnico di Torino}\\
Torino, Italy \\
stefano.dicarlo@polito.it}
}

\maketitle

\begin{abstract}
The automotive domain rapidly increases the embedded amount of complex and interconnected electronics systems. A considerable proportion of them are real-time safety-critical devices and must be protected against cybersecurity attacks.  Recent regulations impose carmakers to protect vehicles against replacing trusted electronic hardware and manipulating the information collected by sensors. Analog sensors are critical elements whose security is now strictly regulated by the new UN R155 recommendation but lacks well-developed and established solutions. This work takes a step forward in this direction, adding integrity and authentication to automotive analog sensors proposing a schema to create analog signatures based on a scrambling mechanism implemented with commercial-of-the-shelf (COTS) operational amplifiers. The proposed architecture implements a hardware secret and a hard-to-invert exponential function to generate a signal’s signature. A prototype of the circuit was implemented and simulated on LTspice. Preliminary results show the feasibility of the proposed schema and provide interesting hints for further developments to increase the robustness of the approach.
\end{abstract}


\section{Introduction}

The automotive domain rapidly increases the amount of hardware and software embedded in a vehicle. The technological evolution of this field, the competition between carmakers, the market’s demands, and the advent of electric cars led to the development of complex and interconnected electronics systems in a vehicle. Many are real-time safety-critical devices and are vulnerable to cybersecurity attacks from two origins \cite{9525579}. The vehicle owners aim to obtain better performance or tamper with annoying features, and professional attackers aim to damage the company’s reputation or take advantage of competitors.

The cybersecurity concern led some authorities to release regulations for vehicles cybersecurity. The Working Party on Automated/Autonomous and Connected Vehicles (GRVA) of the United Nations Economic Commission for Europe (UNECE), through the World Forum for Harmonization of Vehicle Regulations (WP.29), released the UN R155 \cite{UN155} and UN R156 \cite{UN156} regulations. The first regulation requires a management system that focuses on cybersecurity during the entire vehicle lifecycle, and the second establishes requirements for the Software Update Management System. Both are conditions for carmakers to get approval and market access, starting from July 2022. The International Organization for Standardization (ISO) officially released on August 2021 the standard ISO/SAE 21434 \cite{ISO/SAE21434:2021}, which specifies engineering requirements for cybersecurity risk management regarding the concept, development, production, operation, maintenance, and decommissioning of electrical and electronic systems in road vehicles, including their components and interfaces. It also defines a framework that includes requirements for cybersecurity processes and a common language for communicating and managing cybersecurity risk.

The UN R155 regulation forces vehicle manufacturers to identify the critical vehicle elements, perform an exhaustive risk assessment, and treat/manage the identified risks appropriately. Table A1 in this regulation lists a set of vulnerabilities or attack methods related to the corresponding threats. Specifically, items 4.3.7-32 mention a group of physical manipulations of systems that attackers could exploit if not sufficiently protected or hardened. This includes manipulating electronic hardware, replacing authorized electronic hardware, and manipulating the information collected by a sensor.

\begin{figure}[htbp]
\centerline{\includegraphics[width=\columnwidth]{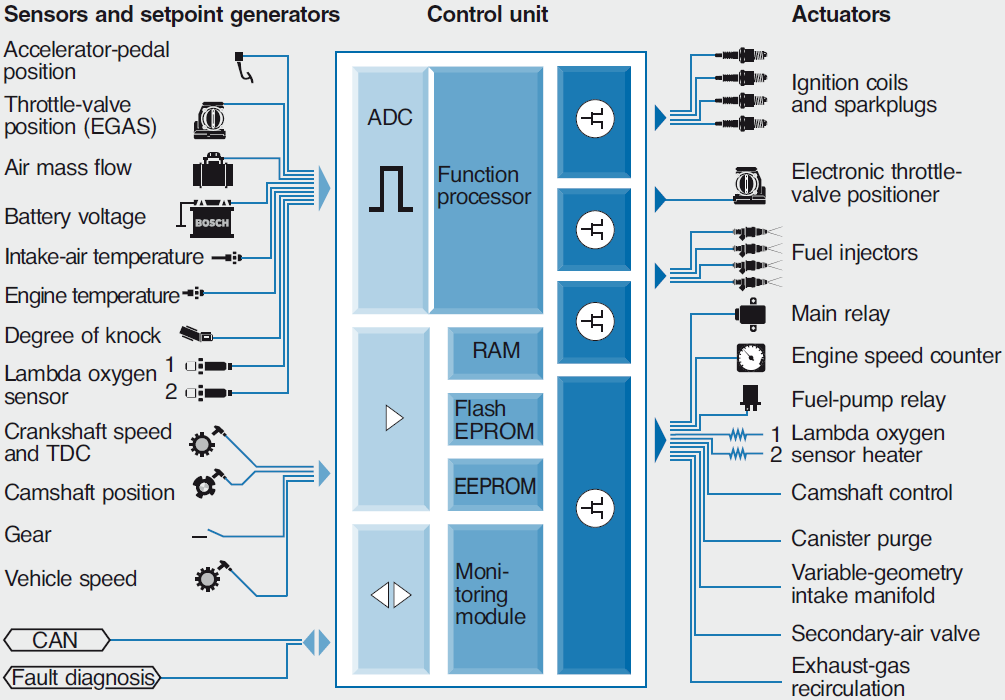}}
\caption{Block diagram of a “Motronic” Engine-Management System ECU. Figure adapted from \cite{gmbh2013bosch}. The figure shows how sensors and actuators represent a crucial element in governing the activities of a vehicle.}
\label{Figure_1}
\end{figure}

The electronic systems in modern vehicles can be separated into several Electronic Control Units (ECUs), handling various subsystems. As an example, Fig.~\ref{Figure_1} shows a “Motronic” Engine-Management System ECU \cite{gmbh2013bosch}. Its heart is a microcontroller. Sensors and setpoint generators produce signals based on the actual physical domain. Actuators triggered by the microcontroller convert the outputs from all functions into mechanical variables.

The most widely used communication protocol between ECUs is the Controller Area Network (CAN) bus, standardized through the ISO 11898 family \cite{ISO11898}. It is a hardware-plus-software protocol that mainly defines how communication happens, how the wiring is configured, and how messages are constructed, thus saving the cost and complexity of “hard-wired” implementations. In the last years, industry and researchers have been trying to improve the CAN bus features against cyber-attacks focusing on network segmentation, encryption, authentication, and intrusion detection \cite{s20082364}.

\begin{figure}[htbp]
\centerline{\includegraphics[width=\columnwidth]{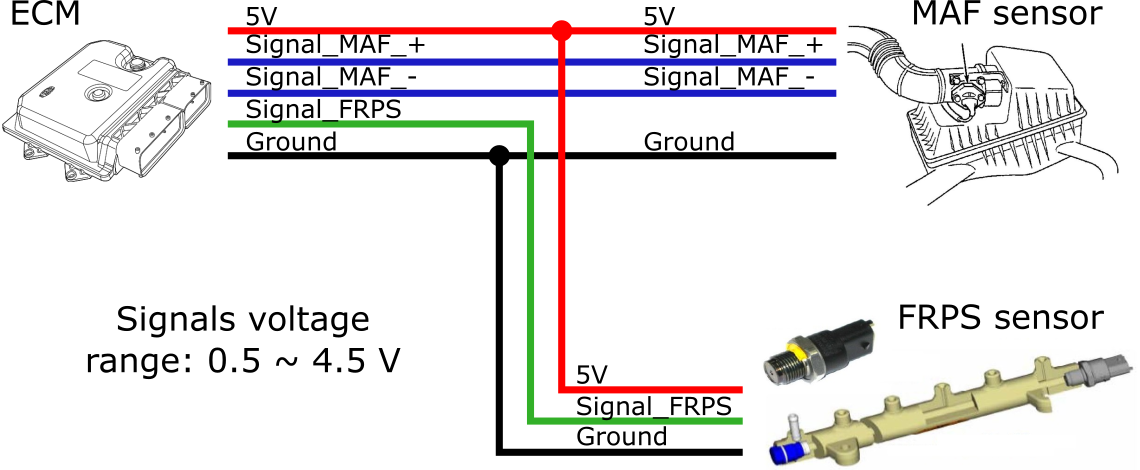}}
\caption{Example of critical sensors connected to the Engine Control Module, one of the most critical ECU governing the activities of the engine in a vehicle.}
\label{Figure_2}
\end{figure}

However, automotive sensors are slightly away from the CAN coverage since this focuses mainly on the communication between ECUs. As commented earlier, the mandatory UN R155 regulation has strict recommendations related to the security of the sensors. Beyond the recommendations, this is important for carmakers to save costs and protect their reputation. Some sensors have a tangible impact on the system and may be of particular interest for attackers, e.g., Fuel Rail Pressure Sensor (FRPS) and the Mass Air Flow (MAF) sensor. They are resistive sensors that generate an analog voltage between 0.5 and 4.5 V. They are connected to an ECU named Engine Control Module (ECM), as shown in Fig.~\ref{Figure_2}, and strongly impact the engine performance. The physical cabling is a vulnerability to the system security since no security feature is implemented.

This vulnerability opens a path to a Man-in-the-Middle (MitM) attack \cite{NistGlossary}. An attacker is positioned between two communicating parties to intercept and alter data traveling between them. In general, MitM attacks have a significant impact on warranty costs. Tampering the vehicle parameters increases vehicle damage risks. In case of damage, the external devices used to mount the attack can be easily removed, making it impossible to prove a tampering action that would lead to a loss of warranty \cite{9486688}.

This paper proposes a novel architecture to secure the communication between analog sensors and the corresponding ECUs in the automotive domain, thus preventing MitM attacks. The architecture exploits two main blocks: an analog hardware secret generator used to create a piece of unique information for every sensor and an analog block able to compute a “hard-to-invert” function exploited to add a signature of the sensor’s signal. The entire architecture has been modeled using LTspice, and electrical simulations show its capability to identify tampering of the signal even in the presence of noise on the sensor’s line.

The paper is organized as follows: Section II introduces the proposed architecture and the main concepts behind its security, while Section III focuses on the electrical implementation. Section IV overviews the experiments performed to validate the architecture, and finally, Section V summarizes the main contributions and outlines suggestions for future improvements.

\section{Proposed Architecture}

A triad of properties formed by confidentiality, integrity, and authenticity is a standard model that forms the basis for developing secure systems. It is used to identify vulnerabilities and methods for addressing security problems and creating effective solutions \cite{samonas2014cia}. Without extreme rigor, it is possible to assume that confidentiality involves keeping data secret or private; integrity ensures the information is trustworthy and free from tampering, and authenticity ensures that the data originates from a trusted source.

In the automotive domain, the CAN bus defines that information from the system should be available in clear text through the On-Board Diagnostic (OBD) port. Therefore, in the proposed architecture, confidentiality is not applicable. However, integrity and authenticity must be preserved at all levels, including analog signals generated by sensors.

In the environment shown in Fig.~\ref{Figure_2}, the target threat is a MitM attack on the ECM connection with the MAF and FRPS sensors. The proposed architecture was designed to comply with the following requirements: (i) be implemented using standard commercial electronic components; (ii) modify as little as possible the design/circuit; (iii) be a low complexity and cost solution; (iv) guarantee a fully analog sensor component. The proposed architecture assumes that the sensor and the security circuit are assembled in the same package. Physical attacks on this package are costly and out of reach for the attacker.

\begin{figure}[htbp]
\centerline{\includegraphics[width=\columnwidth]{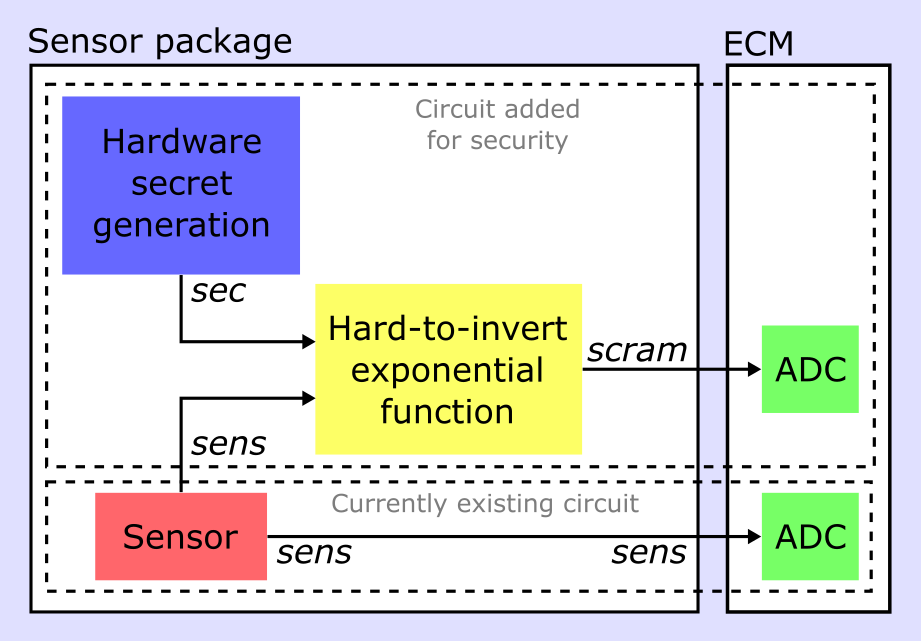}}
\caption{High-level description of the secure analog signal integrity architecture.}
\label{Figure_3}
\end{figure}

Fig.~\ref{Figure_3} shows the proposed architecture that mainly requires modifications at the sensor’s side. The original sensor is enhanced with two additional blocks: a secret generator to create a unique secret for every sensor using an internal randomness source and a hard-to-invert exponential function. The two blocks generate an analog signature required to guarantee integrity and authenticity. From the ECM point of view, a second analog-to-digital converter channel and an additional connection line must be provided.

Following a common practice in the digital domain, the overall idea behind the proposed architecture is to compute an analog signature of the original signal as a one-way function of the signal coupled with a secret, i.e., a function that is easy to calculate on every input, but hard to invert. The concept of ``easy" and ``hard" in the digital domain is related to the computational complexity theory, specifically the theory of polynomial-time problems. In this domain, not being one-to-one is not considered sufficient for a function to be called one-way. However, in the analog domain, the computational capability of an attacker is limited. For this reason, we introduce the concept of hard-to-invert function, i.e., a function that is hard to invert in the analog domain without resorting to powerful digital computation capability.

Equation (1) shows the hard-to-invert exponential function exploited in the proposed architecture. The function produces a scrambled signal (scram) as an exponential value proportional to the secret signal plus the sensor signal (sec+sens).

\begin{equation}
scram \equiv e^{ \propto (sec+sens)}\label{eq}
\end{equation}

Inverting an exponential function requires a logarithm function that is hard to reconstruct given random inputs. This forms one of the fundamental problems in the cryptography field, called the Discrete Logarithm Problem (DLP) \cite{1055638} \cite{10.1145/129902.129904}.

The availability of a secret inside the secure sensor is the key point in this architecture. Discovering the secret means compromising the security of the sensor. An attacker must create a circuit to invert the scrambling operation in the analog domain to achieve this goal. This requires using a logarithm analog circuit to distinguish the secret signal (sec) from the sensor (sens). This second step is hardened by introducing complexity in secret (sec) and sensor (sens) analog processing. Mathematically this is represented by the term \(\propto(sec+sens)\) in (1). It is possible to consider that the cost to implement this attack is high since the attacker must have deep technical knowledge of the exponential function circuit used on the scrambler. The complexity of signals from the analog nature also helps to increase the attack cost. Finally, since every sensor is different, the attack should be repeated on every vehicle.

Both the original sensor’s signal (sens) and the signature signal (scram) are delivered to the ECM that knows the secret (sec) and the used hard-to-invert function. Therefore, the ECU can locally recompute the signature and compare it with the one received from the sensor to verify integrity and authenticity. This verification can be easily performed in the digital domain. Given the possibility of noise in the transmitted signals, the comparison between the original and recomputed signatures takes an error margin into account.

\section{Implementation}

\begin{figure}[htbp]
\centerline{\includegraphics[width=\columnwidth]{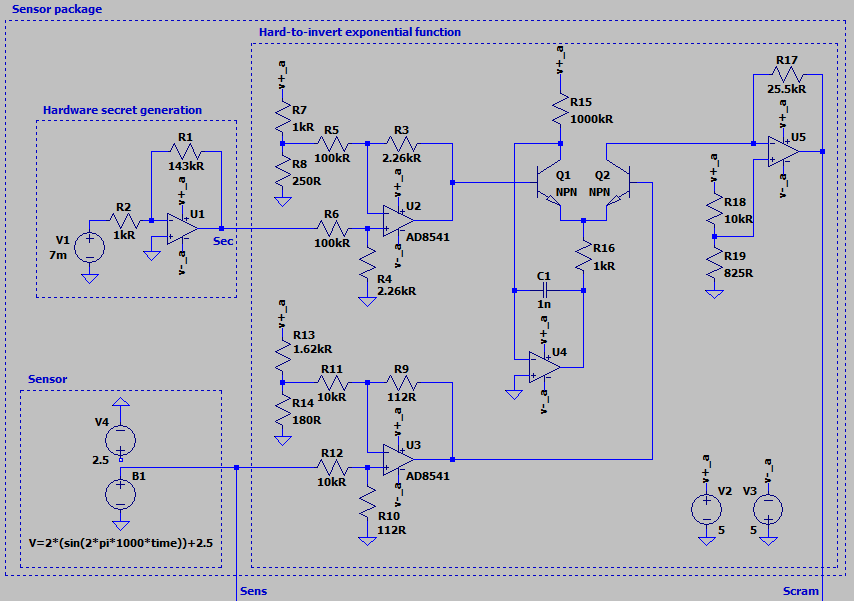}}
\caption{LTspice implementation of the secure sensor architecture. It includes a model of the original sensor, a hardware secret generator, and a hard-to-invert exponential function.}
\label{Figure_4}
\end{figure}

Fig.~\ref{Figure_4} shows an electrical implementation on LTspice \cite{LTspice} of the secure sensor architecture introduced in Fig.~\ref{Figure_3}. In particular, the figure shows the circuit embedded in the sensor package. The package includes the sensor, the hardware secret generation and the hard-to-invert exponential circuit. The entire architecture is based on commercial operational amplifiers. In particular, the Analog Devices AD8541 \cite{AD8541} was selected since it is one of the cheapest operational amplifiers and has high input voltage offset, characteristics desired in this design.

In the automotive domain, sensors have a typical voltage range between 0.5V and 4.5V. This is the voltage range considered for the sensor signal (sens). Since the original (sens) and signature (scram) signals are connected to an ECM’s analog-to-digital converter, the scram signal is designed to work in the same voltage range. Differently, the hardware secret generator produces a signal (sec) arbitrarily defined in the range between -1V and +1V.

The sensor block in Fig.~\ref{Figure_4} has two voltage sources introduced for simulation purposes. V4 is a source to simulate the DC operating point, and B1 is an arbitrary source to generate a sinusoidal 1 kHz 0.5V up to 4.5V signal for transient simulations.

The hardware secret generation is based on an electrical parameter that is random between different devices. Although randomness is necessary between devices, stability over time, temperature, power source variations, and aging is critical in a device. Therefore, the proposed architecture generates the secret using a weak Physically Unclonable Function (PUF) based on manufacturing variations \cite{10.1145/586110.586132}. A secret generated in this way has some advantages: it does not need a battery nor any other permanent power source; it is generated only when required and reliably produces the same result every time \cite{WhitepaperPUF}.

\begin{figure}[htbp]
\centerline{\includegraphics[width=\columnwidth]{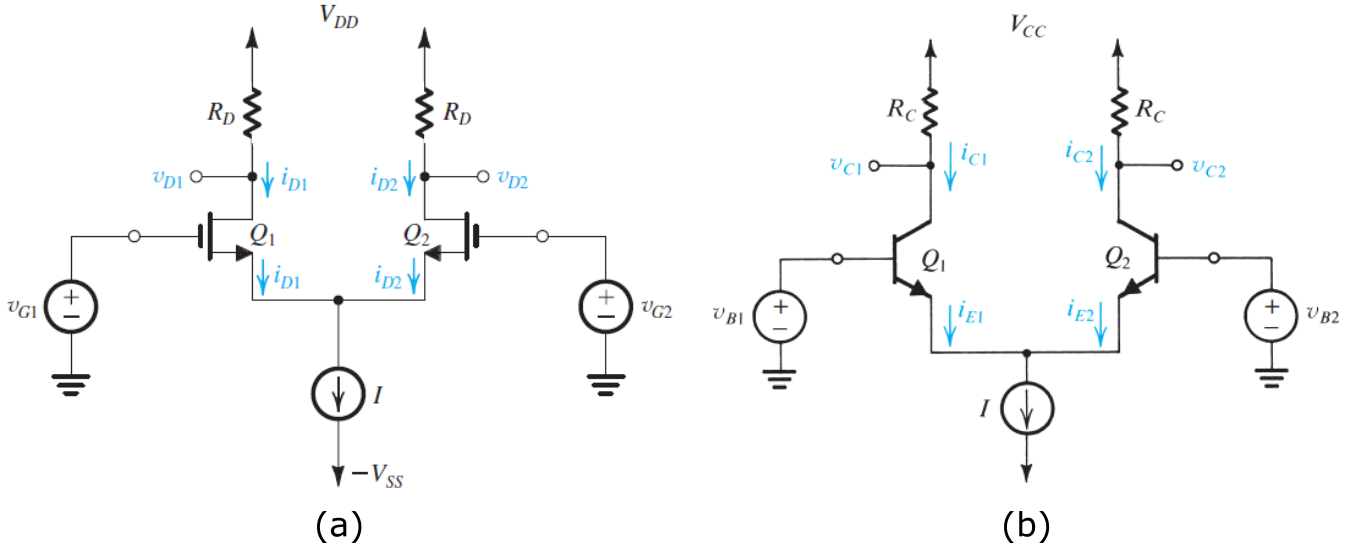}}
\caption{(a) The basic MOS differential-pair configuration. (b) The basic BJT differential-pair configuration [16].}
\label{Figure_5}
\end{figure}

The electrical parameter chosen to generate the secret is the input voltage offset of an operational amplifier. This offset is a result of the internal operational amplifier architecture. The input stage of an operational amplifier is composed of a differential pair, which may be implemented with Metal-Oxide-Semiconductor (MOS) or Bipolar Junction Transistor (BJT) technology (Fig.~\ref{Figure_5}).

The input offset voltage arises due to the unavoidable variations present in this differential pair. In MOS differential pairs, three factors contribute to the offset voltage: variations in load resistances, variations in W/L, and variations in \(V_T\). In Bipolar differential pairs, these factors are from variations in the load resistances, \(\beta\) (from junction area), and other variations on the transistors \cite{sedra2015microelectronic}. The physical sources of these variations can be systematic or random \cite{ArtZirger}. Sources of random variations include edge effects (rough edges), implantation (finite number of charges and distribution), mobility, and oxide effects.

The chosen operational amplifier (AD8541) has a maximum input voltage range of ±7mV. It is used in an amplifier configuration with a gain of 143, resulting in an output voltage range of ±1V. Since the LTspice model of this component has an ideal behavior, the randomness of the offset is emulated by a voltage source in series with the input. This voltage source simulates the voltage offset.

\begin{figure}[htbp]
\centerline{\includegraphics[width=6.33cm, height=4.95cm]{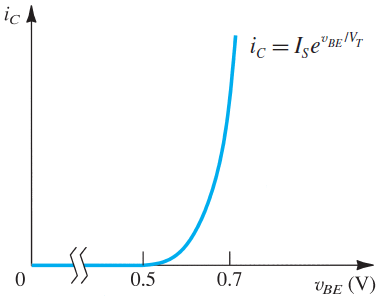}}
\caption{The \(I_C x V_{BE}\) characteristic for an NPN transistor \cite{sedra2015microelectronic}.}
\label{Figure_6}
\end{figure}

The hard-to-invert exponential block implements the exponential function of the \(I_C x V_{BE}\) characteristic of BJT transistors, as shown in Fig.~\ref{Figure_6}. When \(V_{BE}\) is in the range of 0.5V up to 0.7V, the curve is a well characteristic exponential. Thus, this may be considered the core of this block.

\begin{figure}[htbp]
\centerline{\includegraphics[width=\columnwidth]{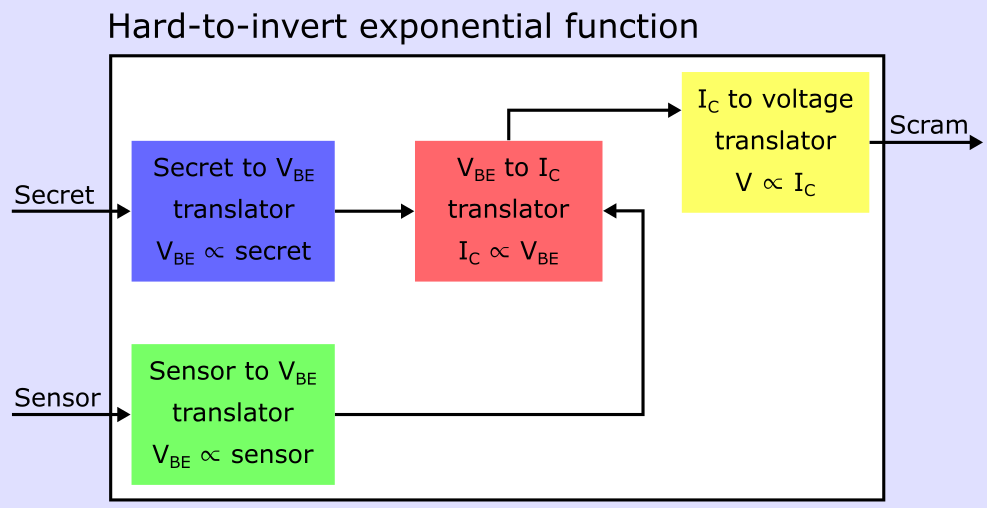}}
\caption{The high-level idea of the hard-to-invert exponential block.}
\label{Figure_7}
\end{figure}

Fig.~\ref{Figure_7} shows a high-level idea of the hard-to-invert exponential block. In the center of the figure is the \(V_{BE}\) to \(I_C\) translator. It comprises two BJT transistors (Q1 and Q2 in Fig.~\ref{Figure_4}) and a current source (U4). Two transistors compensate for the shifting in the \(I_C x V_{BE}\) curve due to temperature effects. The secret and the sensor signals are injected in the level translators U2 and U3, which generate a maximum delta \(V_{BE}\) of 45mV each. When accumulated, secret and sensor signals generate a maximum delta \(V_{BE}\) of 90mV in a range inside 0.5 and 0.7V. This \(V_{BE}\) causes an \(I_C\) variation translated to a voltage variation by the last block (U5).

\begin{figure}[htbp]
\centerline{\includegraphics[width=\columnwidth]{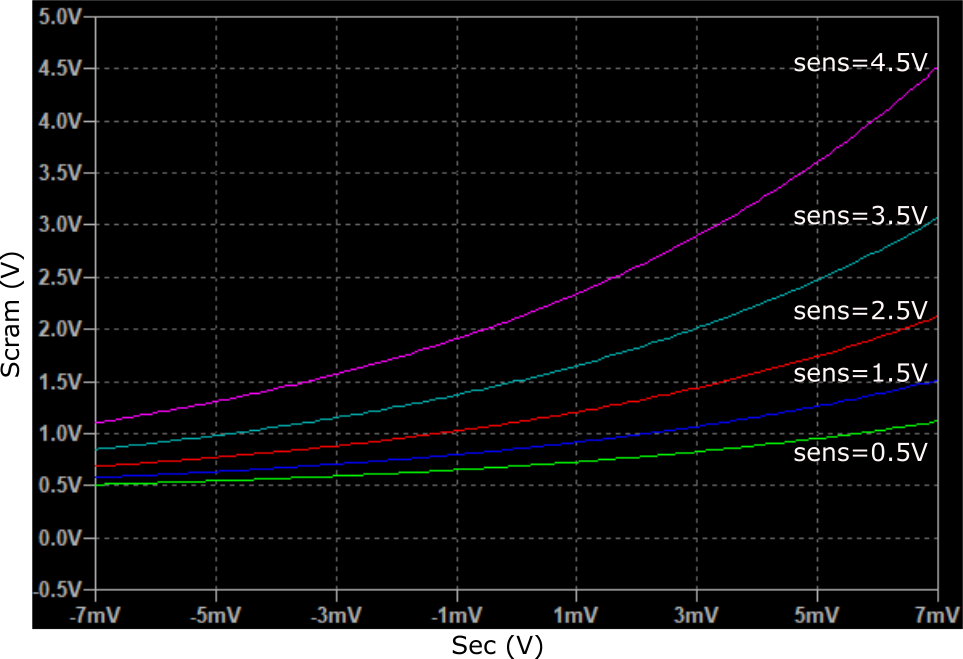}}
\caption{DC simulation of the hard-to-invert exponential function block for different values of the secret and different values of the signal.}
\label{Figure_8}
\end{figure}

Fig.~\ref{Figure_8} shows a DC simulation of this block. According to the stimulus of secret and sensor signals, the output range is between 0.5 and 4.5V.

\section{Results}

This section proposes simulation results showing how the proposed architecture can be tuned for a specific sensor and how the architecture is resilient to attempts of tampering with the sensor signal.

\subsection{Simulation setup and procedures}

\begin{figure}[htbp]
\centerline{\includegraphics[width=\columnwidth]{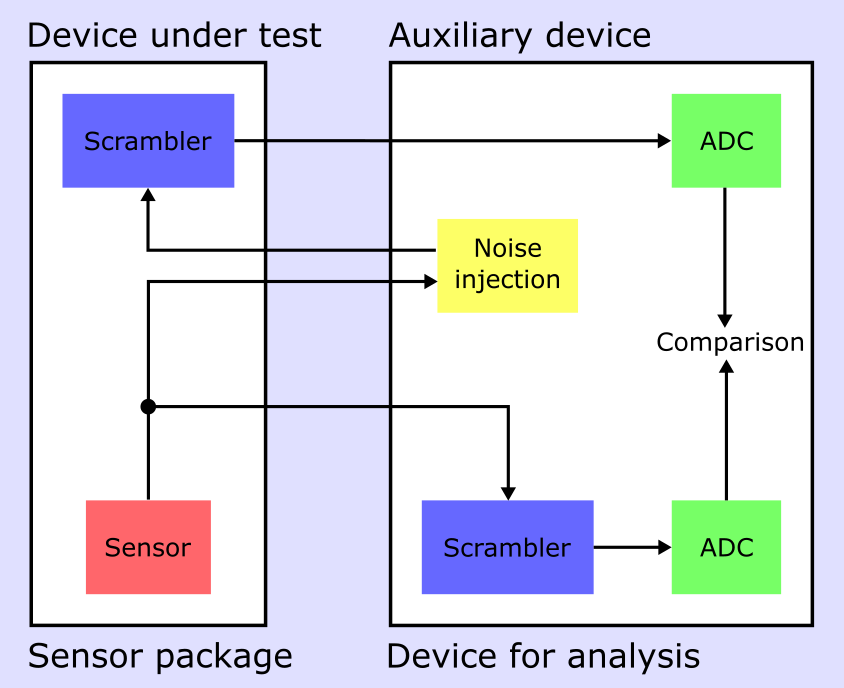}}
\caption{Simulation setup used to analyze the behavior of the secure sensor architecture.}
\label{Figure_9}
\end{figure}

Fig.~\ref{Figure_9} shows the architecture of the simulation setup based on the LTspice. The sensor and the scrambler module, including the hardware secret generator and the hard-to-invert exponential function described in Fig.~\ref{Figure_4}, are connected to an auxiliary device able to emulate the behavior of an ECU and perform an authenticity and integrity check.

Sensors are not ideal devices and are affected by noise. If this noise is not propagating on all paths where the sensor signal is routed in the same way, discrepancies might arise, compromising the proper integrity and authentication checks. A noise is injected in the signal reaching the scrambler block to model this condition while maintaining the other signals unaltered. Typically, sensors in the automotive domain have a 3\% error margin. Therefore, considering that the sensor signal ranges between 0.5 and 4.5V, a 135mV noise (3\% of 4.5V) is considered. In the auxiliary device, a Scrambler circuit sharing the secret of the secure sensor is used to recompute the signature. The transmitted and locally computed signatures are translated in the digital domain and compared. If the difference between the two values is lower than a certain threshold \(V_T\), the authentication and integrity of the sensor are guaranteed. Otherwise, the signal has been tampered with, and a recovery action is required. It is worth mentioning that in a real car, the activities of the auxiliary device are all performed by the ECU performing computations in the digital domain and are implemented here in LTspice to simplify the simulation setup.

\subsection{Comparison threshold \(V_T\) computation}

\begin{figure}[htbp]
\centerline{\includegraphics[width=\columnwidth]{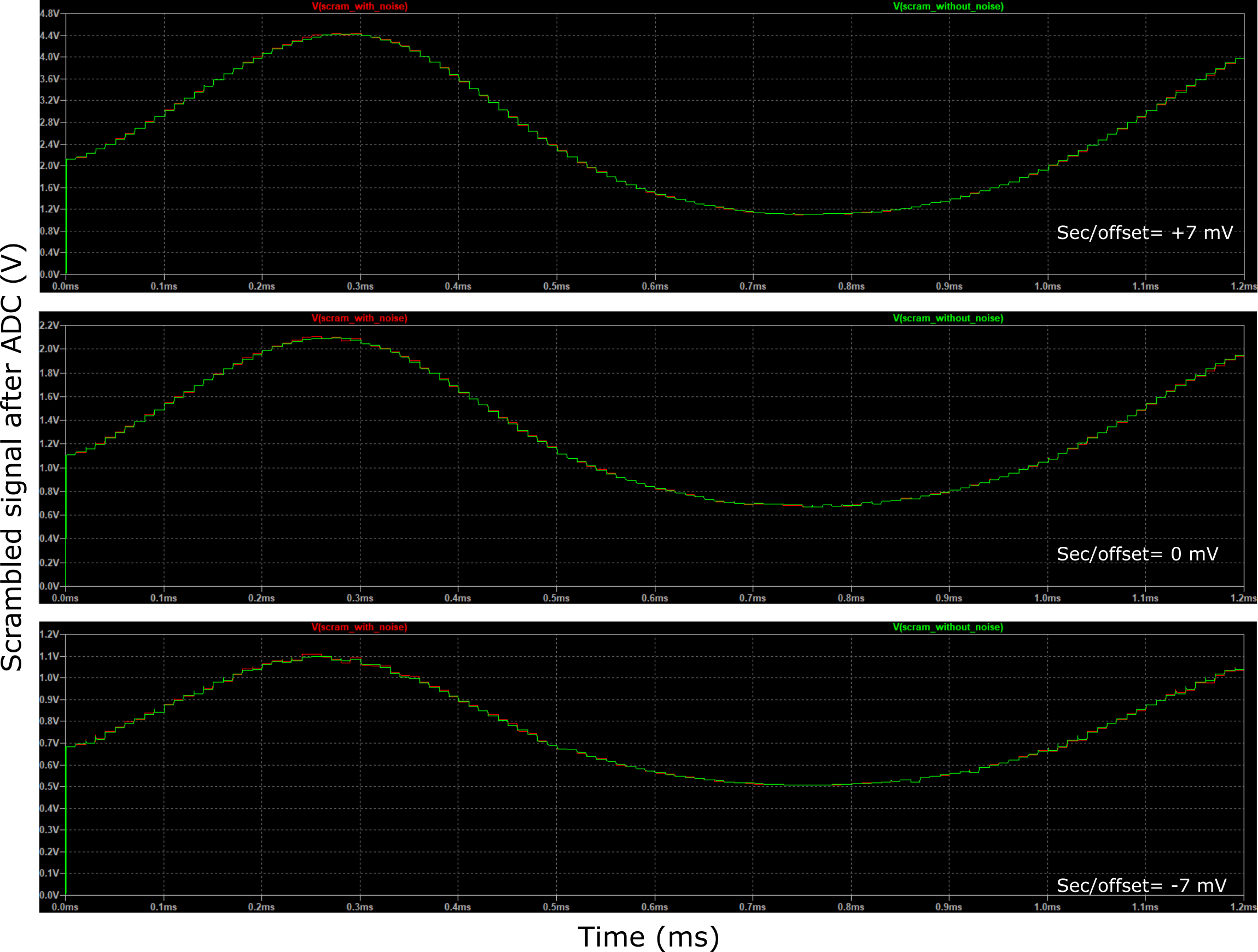}}
\caption{Comparison of the scramble signal generated in the device under test and the scrambler signal computed in the auxiliary device for different secret signals.}
\label{Figure_10}
\end{figure}

The threshold \(V_T\) is the most critical parameter of the proposed architecture. It defines the minimum difference between the transmitted signature and the auxiliary device computed signature detected as an attack. This threshold is a function of the secret implemented in the secure sensor architecture and must be carefully determined for each sensor through characterization. Fig.~\ref{Figure_10} shows the values of the scramble signal generated in the device under test and the one computed in the auxiliary device. The three simulations are obtained with three values of the input range of the hardware secret generator, i.e., -7mV, 0mV, and +7mV, covering the full input range of this signal.

The two signals are different due to the noise injected in the simulation. The simulation outcome is that the worst condition, i.e., maximum difference, is observed for simulation of a secret signal equal to +7 mV. In this case, the difference is 31.586 mV. This value can define the threshold \(V_T\) required to identify potential attacks. It is essential to highlight that this value is small and represents a tiny percentage (i.e., less than 1\%) of the sensor’s dynamic range.

\subsection{Attack detection analysis}

\begin{figure}[htbp]
\centerline{\includegraphics[width=\columnwidth]{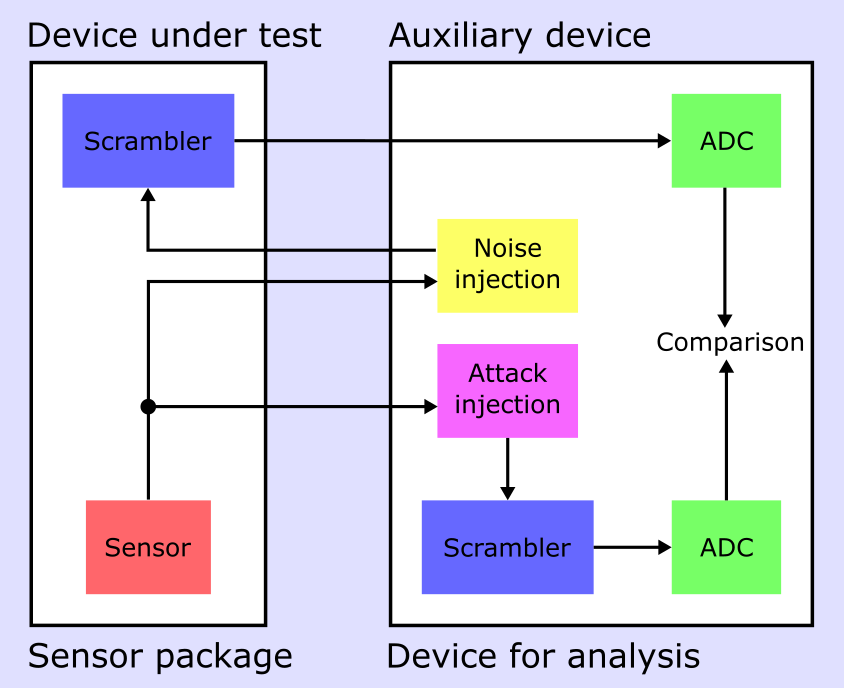}}
\caption{Simulation setup used to emulate the behavior of the MitM attack.}
\label{Figure_11}
\end{figure}

The last part of the simulation analysis aims to study the proposed circuit's ability to detect MitM attacks. Fig.~\ref{Figure_11} shows the simulation setup used to reach this goal. The architecture is similar to the one introduced in Fig.~\ref{Figure_10}, but a malicious user performs an attack by injecting a signal to modify the sensor signal. The evaluation was performed by adding a crescent linear voltage in the sensor signal. As the sensor signal is a 1KHz signal, with a period equal to 1 ms, the attacker increases the injected signal by one mV for every period. Simulations have been performed to identify the minimum injected signal detected as an attack. Simulations were performed using +7 mV as an input signal to the hardware secret generator corresponding to a threshold voltage \(V_T\) equal to 31.586 mV as described before.

\begin{figure}[htbp]
\centerline{\includegraphics[width=\columnwidth]{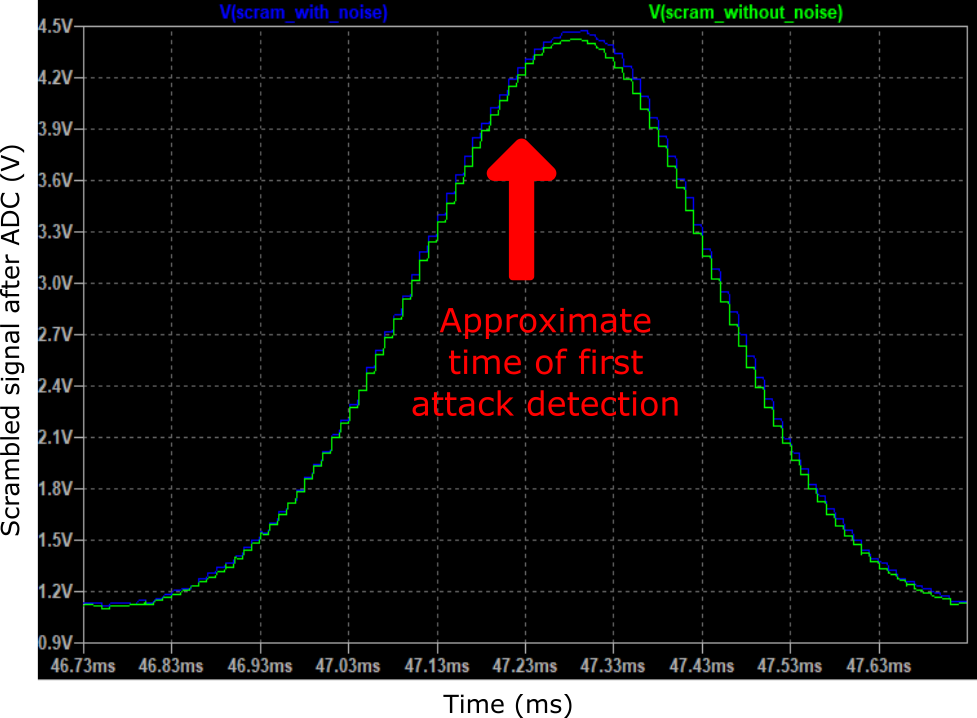}}
\caption{Simulation of an attacker injecting an increasing voltage signal at the output of the sensor.}
\label{Figure_12}
\end{figure}

Fig.~\ref{Figure_12} shows the simulation of the proposed attack. Starting the simulation with a time equal to zero, the difference in the analog-to-digital converters starts increasing while the attacker increases the injected signal and exceeds the threshold of 31.586 mV after approximately 47.23 ms. At this time, the signal injected by the attacker has an amplitude of 47.23 mV. This result shows a reasonable capacity of the architecture proposed and implemented to detect attacks. It is important to remark that the attacker’s goal is to inject an altered sensor signal to modify the behavior of the ECU. To be significant, the modification of the signal must be higher than the 3\% uncertainty of the sensor. Simulations confirm that the architecture is sensitive to variations even lower than this threshold. This is a significant result confirming the potential of this architecture.

\section{Conclusions}

The authenticity and integrity of analog sensors in the automotive domain are mandatory requirements to satisfy increasingly demanding cybersecurity regulations. However, while these properties are nowadays guaranteed in all digital communications in a car, the protection of analog signals is a field still not adequately explored. This paper proposed a first step forward in this direction by offering a new security architecture able to generate an analog sensor signal signature. The architecture generates a secret signal through a weak PUF exploiting the input voltage offset of an operational amplifier. The secret is used to compute a hard-to-invert exponential function to create the signature. Preliminary simulations using LTspice models proved the feasibility of the proposed approach in detecting tampering of the signal. While this represents a significant achievement, additional work is required to move this concept to actual application. Prototyping of the architecture will better assess the stability of the analog signal generation, one of the main critical factors in this architecture. Moreover, future work is required to implement full analog one-way functions that can further increase the security level of the solution.

\bibliographystyle{IEEEtran}
\bibliography{bibliography}

\end{document}